\documentclass[12pt]{article}
\usepackage{graphicx,amsmath,amssymb,url,enumerate,mathrsfs,epsfig,color}
\usepackage{tikz}
\usepackage{float}

\usepackage{amsmath}
\usepackage{amssymb}
\usepackage{yfonts}

\usepackage{tikz,pgfplots}

\usetikzlibrary{calc, patterns,arrows.meta, arrows, shapes.geometric}

\usepackage{graphicx,amsmath,amssymb,url,enumerate,mathrsfs,epsfig,color}

\usetikzlibrary{decorations.text}
\usetikzlibrary{decorations.markings}

\pgfplotsset{compat=1.8}

\usepackage{enumerate}
\usepackage{boxedminipage}
\usepackage{makeidx}
\usepackage{multicol}
\usepackage{xcolor}
\usepackage{mathtools}

\usetikzlibrary{calc, patterns,arrows, shapes.geometric}
\def\QED{\hskip0.1em\hfill\null\ \null\nobreak\hfill
\kern3pt\lower1.8pt\vbox{\hrule\hbox
{\vrule\kern1pt\vbox{\kern1.7pt \hbox{$\scriptstyle
QED$}\kern0.2pt}\kern1pt\vrule}\hrule}}
\title{Large deformations of gradient elastic beams}
\author{Marcelo Epstein\footnote{University of Calgary, Calgary, Canada} and Mohammadjavad Javadi\footnote{Carleton University, Ottawa, Canada}}

\date{}
\begin{document}
\maketitle

\begin{center}
{\it In memory of David W. Murray (1931-2010)}
\end{center}

\begin{abstract}
Budiansky's nonlinear shell theory is particularized to a 2D setting, and thereupon generalized to a fully nonlinear, statically and kinematically exact, theory of strain-gradient elasticity of beams. The governing equations are displayed in their weak and strong forms. A suitable finite element is used to accommodate the new degrees of freedom emanating from the theory and several numerical examples with large geometrical nonlinearities are displayed showing the relative influence of the strain gradient. The numerical apparatus is then applied to permanently magnetized bodies under the action of external magnetic fields.
\end{abstract}

\small{{\bf Keywords}: Shell theory, Geometric nonlinearity, Second-grade elasticity, Magnetized materials

\section{Introduction}

Within the bounds imposed by the classic Kirchhoff-Love hypothesis, Budiansky's nonlinear shell theory \cite{budiansky} is a paradigm of elegance and careful adherence to the underlying continuum mechanics context. Indeed, Budiansky's equations, whether in their weak or strong forms, are both statically and kinematically exact for arbitrary displacements and rotations. When particularized to the theory of large deformations of initially straight plane beams, as done in \cite{1976}, the results are surprisingly clear and amenable to numerical implementation. An important detail, easily checked in the case of beams, is that Budiansky's modified static variables (the axial force and the bending moment in the beam) automatically take into consideration that when a large curvature develops (such as in various Euler elasticas) the neutral axis does not coincide with the centroidal line. Thus, in the case of pure bending, the axial strain at the centroid does not vanish, and it would be wrong, therefore, to conclude that the axial force is proportional to it.  This detail does not need to be incorporated by an ad hoc argument, but rather arises as a natural consequence of the adoption of symmetrized membrane forces in the shell formulation.

The above mentioned features of Budiansky's shell theory, as well as our previous experience with its application to the geometrically nonlinear theory of beams and its numerical implementation, are the driving forces behind this work. Given the renewed interest on micro-scale effects in nano structures, we propose a generalization of the beam formulation presented in \cite{1976} to gradient elasticity. Although in most applications the beam axis is nearly  inextensible even in the presence of very large displacements and rotations, our equations take into consideration the possibility of including both flexural and axial gradient effects.

In fairly recent developments, roboticists have developed soft continuum robots, in contrast to rigid robots, capable of some degree of shape changing and adaptation to different tasks. The combination of highly deformable elastomers and embedded magnetic particles \cite{diller, kim} has already shown the enormous potential for applications of this and other  technologies \cite{hu, pancaldi}. Increasingly sophisticated theoretical models are called for \cite{wang, dadgar} that can capture arbitrarily large displacements and rotations and that incorporate the various scales at play, such as non-local and strain-gradient theories \cite{asghari}. The foundations of the strain-gradient theory can be traced back to \cite{Toupin1962, Toupin1964, mindlin1968}, some of whose main aspects have been incorporated into models of micro/nano beams \cite{Kahro, Asghari2012, dadgar2017}. In the present work we attempt to formulate a fully geometrically nonlinear strain-gradient theory and a model that places no restrictions whatever on the magnitude of displacements and rotations.

Section \ref{sec:linear} offers a brief review of the strain-gradient elasticity  linear beam theory in order to show the expected qualitative and quantitative discrepancies from the classical, non-gradient, results. Beyond the expected stiffening of the response, we also indicate how the strain-gradient theory can in some cases be considered as a singular perturbation of the usual theory. As expected, the fact that the perturbation affects the highest derivative of the governing equation results in the development of boundary layers. Section \ref{sec:nonlinear} is a detailed exposition of the proposed strain-gradient beam theory in the context of arbitrarily large displacements and rotations. Starting from a virtual work expression, the corresponding differential equations and natural boundary conditions are obtained and discussed. The strong formulation is included for completeness only, since the finite-element implementation, presented in Section \ref{sec:finite}, is obtained directly from the principle of virtual work. Fifth-degree polynomials are used as shape functions, employed also by \cite{dadgar2017}. The implementation is based on the idea that, in principle at least, all that needs to be coded is the expression of the internal (and external) virtual work in the typical element. The formulation being purely Lagrangian, everything else is left to the Newton-Raphson method, which automatically produces the required residuals and Jacobian matrices by systematically calling the virtual work routine. This strategy may not be the most economical one to pursue, but has the merit of being easily coded and avoiding many of the paraphernalia encountered in the linear finite-element theory. The code is validated in Section \ref{sec:examples} under very severe deformation regimes. An important application to the deformation of magnetized beams subjected to an external magnetic field is included and, in this context, the numerical algorithm reveals the possible existence of multiple solutions.

\section{Higher-order effects in elastic beams}
\label{sec:linear}
The purpose of this section is to review the influence of higher-order elasticity within the simplest possible context, namely, the linear (small deflection) theory. We will, accordingly, assume that  the internal virtual work of the beam is given by the expression
\begin{equation} \label{eq1}
IVW=\int\limits_0^L( EI w'' \delta w''+B w''' \delta w''') \,  dx .
\end{equation}
In this equation $x$ is the running coordinate, $L$ is the beam length, $w=w(x)$ is the transverse displacement, primes denote $x$-derivatives, $EI$ is the flexural constant, and $B$ is a positive material constant attributable to higher-order effects, which in this model affect the third derivative of the displacement (the derivative of the curvature).

For definiteness, we consider a cantilever beam fixed at $x=0$ with a vertical force $P$ and/or a couple $M$ applied at the free end. The virtual work of these loads is given by
\begin{equation} \label{eq2}
EVW=P \delta w(L) + M \delta w'(L).
\end{equation}

By the principle of virtual work, we obtain the following differential equation
\begin{equation} \label{eq3}
EI w^{iv}-B w^{vi}=0,
\end{equation}
with the boundary conditions
\begin{equation} \label{eq4}
\left\{
\begin{matrix}
w(0)=0\\w'(0)=0\\ w'''(0)=0 \\ w'''(L)=0 \\ EIw''(L)-Bw^{iv}(L)=M \\
B w^v (L) = P
\end{matrix}
\right.
\end{equation}

We can already see something surprising, in the sense that the classical shear force at the end of the beam is absorbed by the fifth (rather than the third)  derivative of the deformation. Clearly, for very small values of $B>0$ we can regard this problem as one of a singular perturbation. For vanishing $B$, the differential equation reduces its order, so that the additional boundary conditions are satisfied at the expense of a boundary layer.

 Adopting the non-dimensional axial variable $\xi=x/L$, and (without danger of confusion) letting primes denote derivatives with respect to  $\xi$, we obtain
\begin{equation} \label{eq5}
w''''-\beta w^{vi}=0,
\end{equation}
with $\beta=B/(EIL^2)$. The corresponding boundary conditions are
\begin{equation} \label{eq6}
\left\{
\begin{matrix}
w(0)=0\\w'(0)=0\\ w'''(0)=0 \\ w'''(1)=0 \\ w''(1)-\beta w^{iv}(1)=ML^2/(EI) \\
\beta  w^v (1) = PL^3/(EI)
\end{matrix}
\right.
\end{equation}

Consider the case $P=0$. The solution is given by
\begin{equation} \label{eq7}
w=\frac{M L^2\xi^2}{2EI}.
\end{equation}
For pure bending, therefore, the higher-order terms have absolutely no bearing on the solution, which coincides with the classical beam solution.

For the case $M=0$ and $P\ne 0$, on the other hand, we obtain
\begin{equation} \label{eq8}
w=\left(\frac{\beta^{3/2}\left(1- e^{-\xi/\sqrt{\beta}}\right)\left(e^{1/\sqrt{\beta}}+e^{\xi/\sqrt{\beta}}\right)}{1+e^{1/\sqrt{\beta}}}-\beta \xi +\frac{\xi^2}{6}(3-\xi)\right)\frac{PL^3}{EI}
\end{equation}

Figure \ref{fig1} shows the influence of the parameter $\beta$ on the deflection curve of the beam. For $\beta=0$ we recover the classical formula. Notice that for values $\beta \ge 1$ the curve remains very close to a final limiting shape. In fact, it is not difficult to prove that the limit curve as $\beta \to \infty$ is given by the quadratic equation
\begin{equation} \label{eq9}
w_{\beta \to \infty}=\frac{1}{4} \xi^2\,\frac{PL^3}{EI}.
\end{equation}

\begin{figure}[H]
	\centering
\includegraphics[scale=0.8, trim=0.in 0in 0in 0.0in, clip]{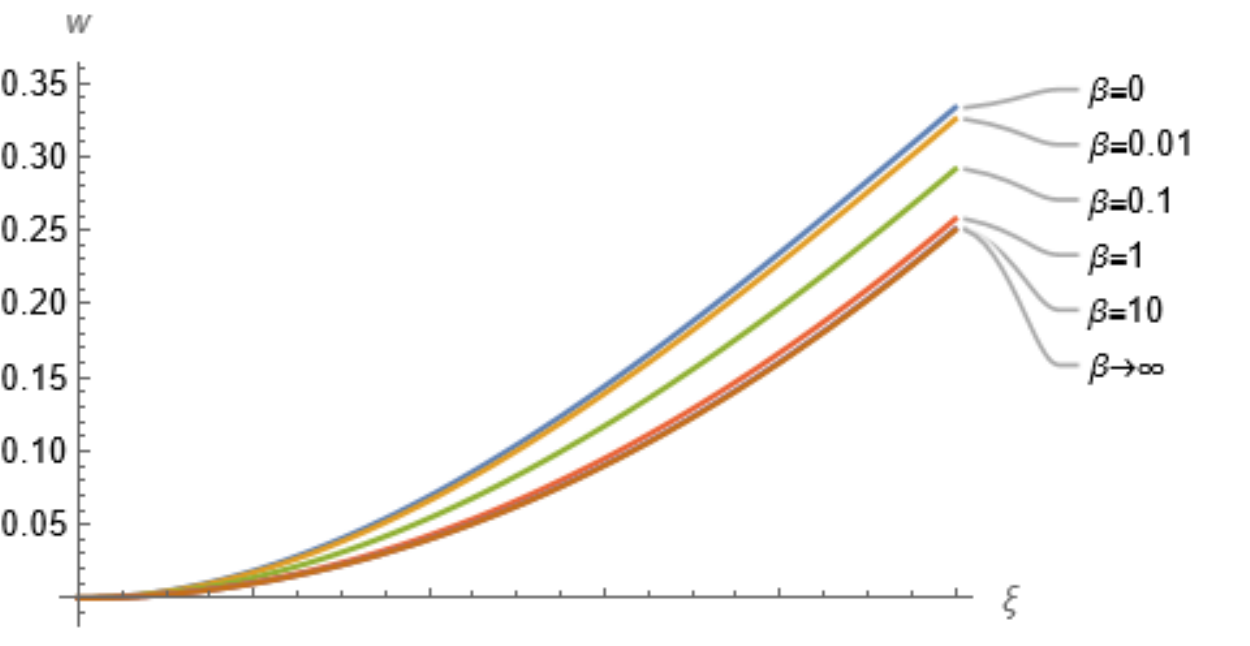}
\caption{Beam deflection for various values of $\beta$ with $P=EI/L^3$}\label{fig1}
\end{figure}

As already pointed out, for very small values of $\beta$ the problem can be regarded as one of a singular perturbation of the regular strain theory. This feature is demonstrated in Figure \ref{fig2}, where the third derivative of the deflection is ploted for two values of $\beta$. The boundary-layer effect is apparent. For the ordinary beam theory, the third derivative is equal to a constant.

\begin{figure}[H]
	\centering
\includegraphics[scale=0.8, trim=0.in 0in 0in 0.0in, clip]{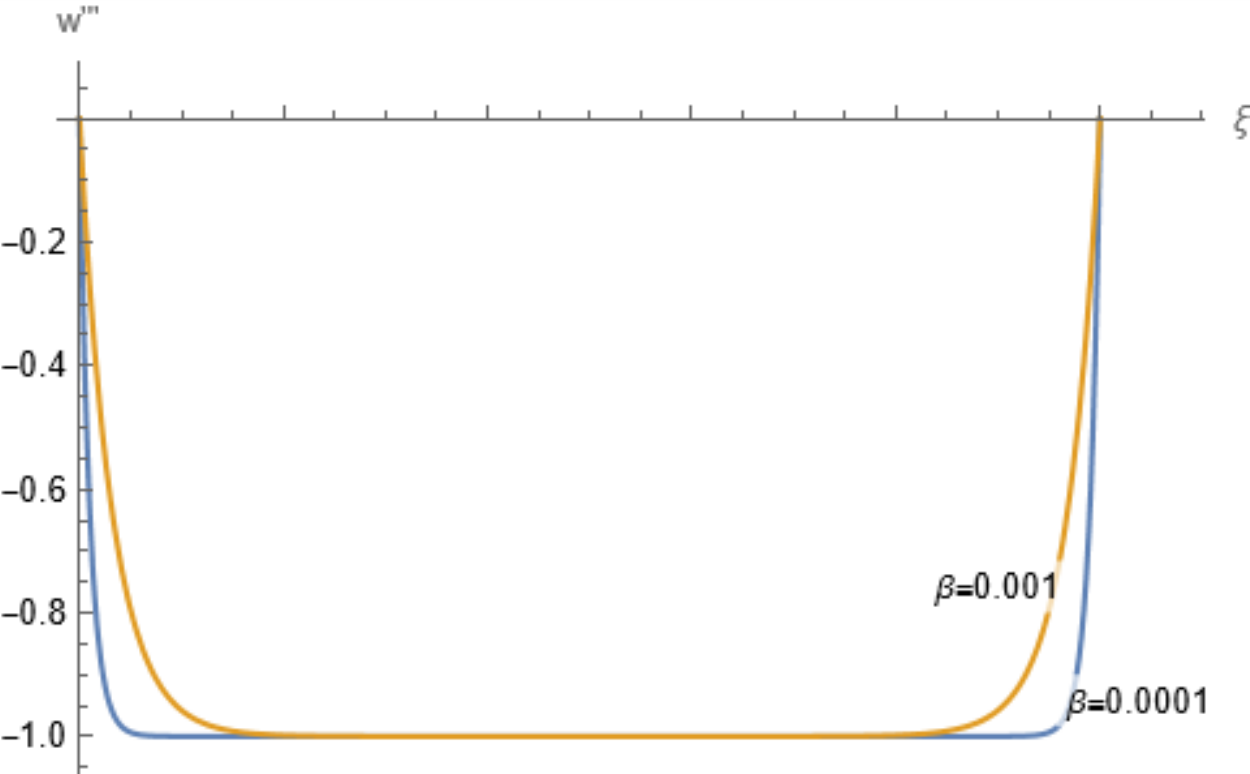}
\caption{The boundary-layer effect}\label{fig2}
\end{figure} 

At this point it is worth remarking that, although the strain-gradient theory introduces an internal length scale, it does not necessarily agree with the predictions of a legitimately non-local theory, such as that pioneered by Eringen decades ago. On the contrary, the non-local theory predicts a softening of the material response. In this regard, see \cite{sudak, xu, xu1}. Negative values of $\beta$ lead to instability. We note that possibly negative energy terms involving an interaction between the strain and its gradient are possible, as cleverly  discussed in \cite{maranganti} in regard to resolving a dispersion paradox. These extra terms, however, do not seem to affect the qualitative character of the static results.

\section{The geometrically nonlinear theory} 
\label{sec:nonlinear}

A (plane) beam will be regarded as an (initially straight) segment $0\le x \le L$ whose deformation in the plane $x,z$ is described by the respective components $u(x)$ and $w(x)$ of the displacement vector. Denoting by primes the derivatives with respect to the referential axial coordinate $x$, the axial (Lagrangian) stretch is obtained (see, e.g. \cite{1976}) as
\begin{equation} \label{eq10}
e= u'+\frac{1}{2}\left( u'^2+w'^2\right).
\end{equation}
The curvature of the deformed beam axis is similarly obtained as
\begin{equation} \label{eq11}
\kappa=\frac{w''(1+u')-w'u''}{(1+2e)^{3/2}}.
\end{equation}
These quantities, $e$ and $\kappa$, can be regarded as exact measures of strain for a theory of beams in the absence of strain-gradient effects. Alternatively, we may replace the curvature measure by the quantity
\begin{equation} \label{eq12}
\chi = w''(1+u')-w'u''.
\end{equation}
We note that for small axial strains ($e << 1$) $\chi$ and $\kappa$ become indistinguishable, even if the derivatives of the displacements are very large. On the other hand, the measures $e$ and $\chi$ can be used also for large axial strains, since any given consitutive equation that can be expressed in terms of $e$ and $\kappa$ can also be expressed in terms of $e$ and $\chi$. This argument was famously advocated by Budiansky (see, e.g., \cite{budiansky}). It has the advantage of involving only rational expressions of the kinematic variables, without sacrificing exactness.

In the absence of of strain-gradient effects, the contribution of the internal stresses to the variational formulation of the theory can be encapsulated in the internal virtual work expression
\begin{equation} \label{eq13}
IVW=\int\limits_0^L (N \delta e + M \delta \chi) \, dx.
\end{equation}
In this expression, $N$ and $M$ are, respectively, the axial force and the bending moment along the beam axis. Although this expression can be justified from an approach that, starting from a three-dimensional formulation, enforces the classic assumption (cross sections remain rigid and perpendicular to the axis), in the intrinsic approach adopted here this justification is unnecessary for the derivation of the governing equations \cite{1976}.

In the strain gradient theory, the expression for the internal virtual work is assumed to involve also the first derivatives of $e$ and $\chi$. Thus
\begin{equation} \label{eq14}
IVW=\int\limits_0^L (N \delta e + M \delta \chi + K \delta e'  + H \delta \chi')\,dx,
\end{equation}
where $K$ and $H$ are interpreted as new internal force resultants associated, respectively, with the axial and flexural strain measures. 

Naturally, one expects that the external virtual work may also incorporate new kinds of external mechanical agents beyond ordinary forces and couples (whether concentrated or distributed). Nevertheless, unless a cogent argument can be advanced for the presence and the mechanism of actuation of these new external agents, it is prudent to assume that they vanish identically along the beam and/or at the boundaries (the two end points). Similarly, conditions of support associated with higher derivatives of the classical degrees of freedom can be conceived (such as a support that prescribes the vanishing of the second or third derivatives of the displacements). One of the merits of the weak formulation (in our case, the principle of virtual work) is that it provides the so-called natural boundary conditions establishing precisely what kinds of supports or forces can be prescribed in a manner that is consistent with the internal virtual work expression. This merit is also preserved in the corresponding numerical implementation of the weak formulation, that is, in the finite element method.

Consider, therefore, the case of a beam loaded only with distributed forces $h(x)$ and $q(x)$ in the axial and transversal directions, respectively. Assume, moreover, that these are ``dead loads'', in the sense that they preserve their magnitude and direction during the process of deformation. At this point we are not specifying the boundary conditions so as to let the principle of virtual work provide us with the available options. In the static case, this principle establishes that
\begin{equation} \label{eq15}
\int\limits_0^L (N \delta e + M \delta \chi + K \delta e'  + H \delta \chi')\,dx \equiv \int\limits_0^L ( h \delta u + q \delta w) \, dx +EVW_{boundary},
\end{equation}
where the last term represents the virtual work of the applied boundary forces and couples, if any. We will only admit as external mechanical agents at the boundaries ordinary forces and couples. The virtual work of a couple $C$ acting at a point $P$ of the beam axis is given by $C \delta\theta_P$. The rotation $\theta_P$ is expressed as
\begin{equation} \label{eq15a}
\theta_P=\arctan\left(\frac{w'}{1+u'} \right)_P ,
\end{equation}
as shown in Figure \ref{fig:rotation}.
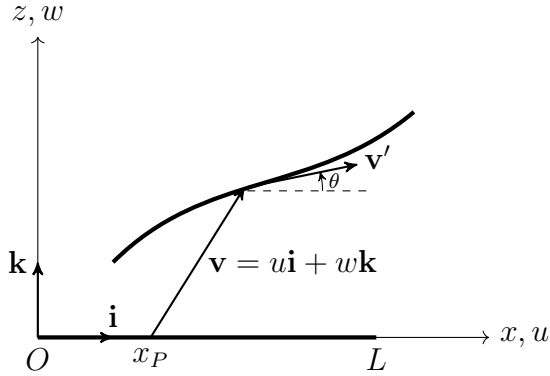
\begin{figure}[H]
\begin{center}
\begin{tikzpicture} [scale=1.0]
\draw[->] (0,0)--(6,0);
\node[right]  at (6,0) {$x,u$};
\draw[->] (0,0)--(0,4);
\node[above] at (0,4) {$z,w$};
\draw[ultra thick,] (0,0)--(4.5,0);
\node[below] at (4.5,0) {$L$};
\node[below] at (0,0) {$O$};
\draw[ultra thick] (1,1) to [out=45, in=220] (5,3);
\draw[thick, -stealth'] (1.5,0)--(2.75, 2.);
\node[right] at (2.12, 1) {${\bf v}=u{\bf i}+w{\bf k}$};
\node[below] at (1.5,0) {$x_P$};
\draw[thick, -stealth'] (0,0)--(0,1);
\draw[thick,-stealth'] (0,0)--(1,0);
\node[above] at (1,0) {$\bf i$};
\node[left] at (0,1) {$\bf k$};
\draw[thick,-stealth'] (2.785,2)--(4.25,2.3);
\node[right] at (4.2,2.4) {${\bf v}'$};
\draw[dashed] ((2.785,1.95)--(4.4,1.95);
\node[above] at (3.95,1.85) {$_\theta$};
\draw [-stealth',domain=0:15] plot({2.785+1*cos(\x)},{1.95+1.*sin(\x)});
\end{tikzpicture}
\end{center}
\caption{Displacement vector  $\bf v$ and its derivative}
\label{fig:rotation}
\end{figure}

If enough differentiability is prescribed, repeated integration by parts of the left-hand side of Equation (\ref{eq15}) yields the differential equations
\begin{equation} \label{eq16}
-\left(N(1+u')\right)' -(Mw'')'-(Mw')''+\left(K(1+u')\right)''-(Ku'')'-(Hw''')'+(Hw')''' = h,
\end{equation}
\begin{equation} \label{eq17}
-(Nw')'+\left(M(1+u')\right)''+(Mu'')'+(Kw')''-(Kw'')'-\left(H(1+u')\right)'''+(Hu''')'=q,
\end{equation}
with the following boundary conditions to be prescribed at $x=0$ and at $x=L$
\begin{equation} \label{eq18}
{\rm either~} u {\rm~or~}N(1+u')+Mw''+Ku''+Hw'''-(Mw')'-(K(1+u'))'-(Hw')'' ,
\end{equation}
\begin{equation} \label{eq19}
{\rm either~} w {\rm~or~} Nw'-Mu''+Kw''-Hu'''-(M(1+u'))'-(Kw')'+(H(1+u'))'',
\end{equation}
\begin{equation} \label{eq20}
{\rm either~} u' {\rm~or~}-Mw'+K(1+u')+(Hw')',
\end{equation}
\begin{equation} \label{eq21}
{\rm either~} w' {\rm~or~}M(1+u')+Kw'-(H(1+u'))' ,
\end{equation}
\begin{equation} \label{eq22}
{\rm either~} u'' {\rm~or~}Hw' ,
\end{equation}
\begin{equation} \label{eq23}
{\rm either~} w'' {\rm~or~}H(1+u') .
\end{equation}
We observe that, under the assumption that there is no mechanism to prescribe the second derivatives of the displacements at a boundary point, to satisfy Equations (\ref{eq22}) and (\ref{eq23}) on the boundaries we must stipulate $H=0$. Equations (\ref{eq20}) and (\ref{eq21}) can be projected onto the tangential and normal directions to the deformed beam, which leads to the following equivalent alternatives at the boundaries
\begin{equation} \label{eq24}
{\rm either~} \delta e =0 {\rm~or~} K {\rm~ is ~prescribed},
\end{equation}
and
\begin{equation} \label{eq25}
{\rm either~} \delta \theta =0 {\rm~or~a~couple~} C {\rm~ is ~prescribed}.
\end{equation}
Since there is no simple support mechanism to prescribe the elongation $e$ at a boundary, the alternative (\ref{eq24}) must be satisfied by stipulating $K=0$ at the boundaries. The alternative (\ref{eq25}) expresses the classical boundary condition given by a clamp or, in its absence, by the application of a (possibly vanishing) boundary couple $C$. Finally, Equations (\ref{eq18}) and (\ref{eq19}) correspond to the specification of a horizontal and/or a vertical force, respectively. These natural boundary conditions are automatically implied in a Galerkin-based finite element implementation.

The consitutive equations will be assumed of the form
\begin{equation} \label{eq26}
\begin{split}
N&=EAe\\
M&=EI\chi\\
K&=Ce'\\
H&=B\chi',
\end{split}
\end{equation}
where $EI$ and $B$ are the flexural constants already used in Section \ref{sec:linear}, and $EA$ and $C$ are material constants associated with the axial strain and its derivative.  The linearity between stresses and corresponding strains is in no way an impediment for the development of large deflections and rotations. Thus, the theory is kinematically exact. Further material nonlinearities can be introduced by replacing these simple constitutive equations with any other expressions without altering the rest of the formulation.

\section{The finite-element formulation}
\label{sec:finite}

The core of a finite-element implementation is a routine that codes the internal virtual work in the typical element. All the rest, including the evaluation of the residuals and the construction of the Jacobian matrix that drives the Newton-Raphson iterations, can be boiled down to repeated calls to this basic routine. Naturally, there are many possible shortcuts (such as the avoidance of the calculation of Jacobian entries that are known a priori to be zero), but these issues do not concern us here, particularly within the context of a one-dimensional framework, even when invoving unlimited large deflections and rotations. To account for the more sophisticated nature of the theory, the finite element used in \cite{1976} will be enlarged to include two additional shape functions. If in \cite{1976} 3rd degree polynomials were found to be sufficiently accurate for even extreme conditions, the interpolating polynomials will be now of degree 5. These shape functions will be used for both displacement components. The associated nodal coefficients, therefore, will carry the physical meaning of displacements, slopes and ``curvatures'' (or rather second derivatives of the displacements) at each node of the element.\footnote{Within the domain of the geometrically linear theory, an alternative set of shape functions was proposed in \cite{asghari}, where more involved shape functions are used that capture the exact shape of an unloaded single element in the geometrically linear theory.} The beam is assumed to be initially straight.

For a typical element of length $h$, the 6 shape functions in terms of the local coordinate $\xi=x/h$ are
\begin{equation} \label{eq50}
\begin{split}
N_1&=-(\xi-1)^3 (1+3\xi+6\xi^2) \\
N_2&=-(\xi-1)^3 \xi (1+3\xi) h\\
N_3&=-(\xi-1)^3 \xi^2 h^2/2\\
N_4&=\xi^3(10-15\xi+6\xi^2)\\
N_5&=\xi^3(-4+7\xi-3\xi^2)h\\
N_6&=(\xi-1)^2 \xi^3 h^2/2
\end{split}
\end{equation}
The different dimensionality of the various shape functions is introduced to preserve the physical dimensionality of the nodal variables. Indeed, at each node $I$ the 6 degrees of freedom are ordered as $u_I,u'_I,u''_I,w_I,w'_I,w''_I$. The interpolated displacement fields for the element comprised between the successive nodes $I$ and $I+1$ are, accordingly, given by
\begin{equation} \label{eq51}
\begin{split}
u&=u_I N_1+u'_I N_2+u''_IN_3+u_{I+1} N_4+u'_{I+1} N_5 +u''_{I+1} N_6 \\
w&=w_I N_1+w'_I N_2+w''_IN_3+w_{I+1} N_4+w'_{I+1} N_5 +w''_{I+1} N_6
\end{split}
\end{equation}
The corresponding contribution of the element to the internal virtual work is obtained by introducing these expressions into Equation (\ref{eq14}) and using Equations (\ref{eq10}) and (\ref{eq12}) and the constitutive equations (\ref{eq26}). Taking into account that all the equations entailed in this process are polynomial expressions in the kinematic variables and their derivatives, it follows that all the integrations involved can be carried out exactly (either analytically or by means of the correspondingly exact Gauss integration).

Assume that the beam has been divided into $n$ elements (of equal or unequal lengths) and that the nodes are numbered successively starting from node 0 at the left end of the beam. The discretized Galerkin equilibrium equations can be thought of as the result of successively setting each of the nodal variations to 1 while the others are made to vanish. Thus, given an initial guess for the solution, namely a collection of the $6(n+1)$ nodal variables (satisfying, of course, the displacement boundary conditions), the (un-balanced) residual corresponding to the equation number $i$ is obtained by setting all variations to zero except the variation corresponding to one of the degrees of freedom at the node which is the whole part of the quotient $(i-1)/6$. The degree of freedom of interest at that node is equal to 1 plus the remainder of this division. For example, the residual of the equation number 14 corresponds to the second degreee of freedom of node number 2. Thus, we set all the variations to zero except $\delta u_2'=1$ and evaluate the difference between the internal and the external virtual works. Clearly, this involves only elements 2 and 3, so the number of operations is minimal.

To proceed with the Newton-Raphson method, we need to evaluate the Jacobian matrix of the equilibrium equations at the given guess. Since, as already pointed out, all these equations are polynomials (at most of the 8th degree) in all the nodal variables, the entries of this matrix can be obtained exactly, if so desired, by means of Lagrangian interpolation. Alternatively, approximate entries can be obtained by simple first or second order forward or central differentiation. Again, this can be achieved, by keeping the variation at its unit value, but changing successively by a small amount the value of each of the nodal variables. In short, as already remarked, every part of the procedure consists of repeated calls to the element virtual work routine.\footnote{In this respect it is unproductive to use the terminology and notation of the standard linear finite element formulations.}

\section{Numerical examples}
\label{sec:examples}

\subsection{Pure bending}

To demonstrate the accuracy of the equations and of the finite element implementation we consider first the case of pure bending of a cantilever beam subjected to a couple that deforms it into a full circle. To simulate axial inextensibility we adopt, in some system of units, a beam of unit length ($L=1$) with very large cross-sectional area ($A=100000$) with $C=0$. The bending constant is $EI=1$. The solution should be independent of the ratio $\beta=B/(EIL^2)$. The applied couple theoretically necessary to bend the beam into a full circle is $2\pi EI/L$, which we apply to the free end of a cantilever. We divide the beam into a mere 4 finite elements of equal length. The numerically obtained results are displayed in Figure \ref{fig4} for $\beta=0$, and in Figure \ref{fig5} for $\beta=1$. The graphs are practically identical, as evidence of the fact that under pure bending the strain gradient is inoperative.
\begin{figure}[H]
	\centering
\includegraphics[scale=0.8, trim=0.in 0in 0in 0.0in, clip]{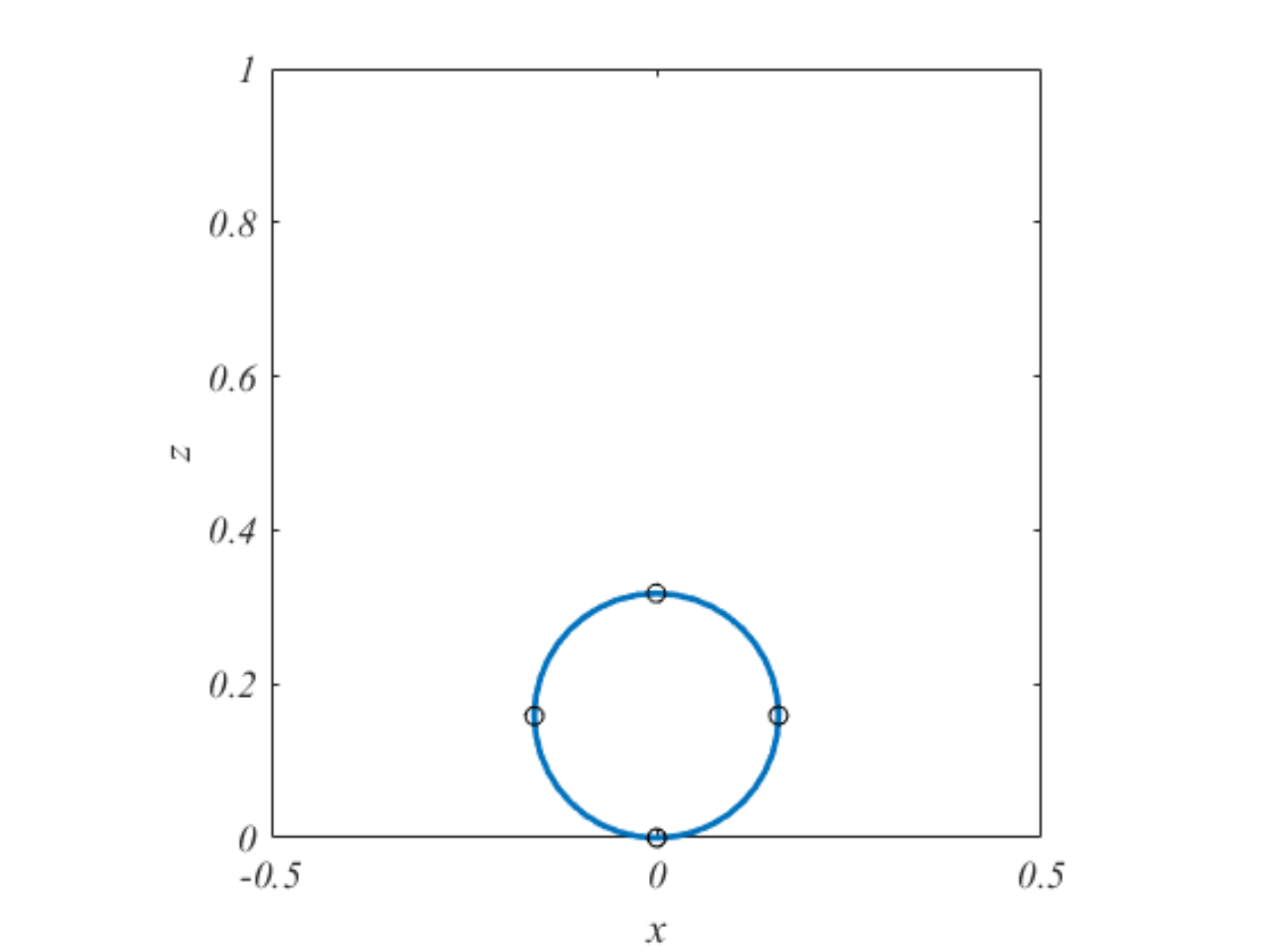}
\caption{Pure bending with $\beta=0$}\label{fig4}
\end{figure} 

 \begin{figure}[H]
	\centering
\includegraphics[scale=0.8, trim=0.in 0in 0in 0.0in, clip]{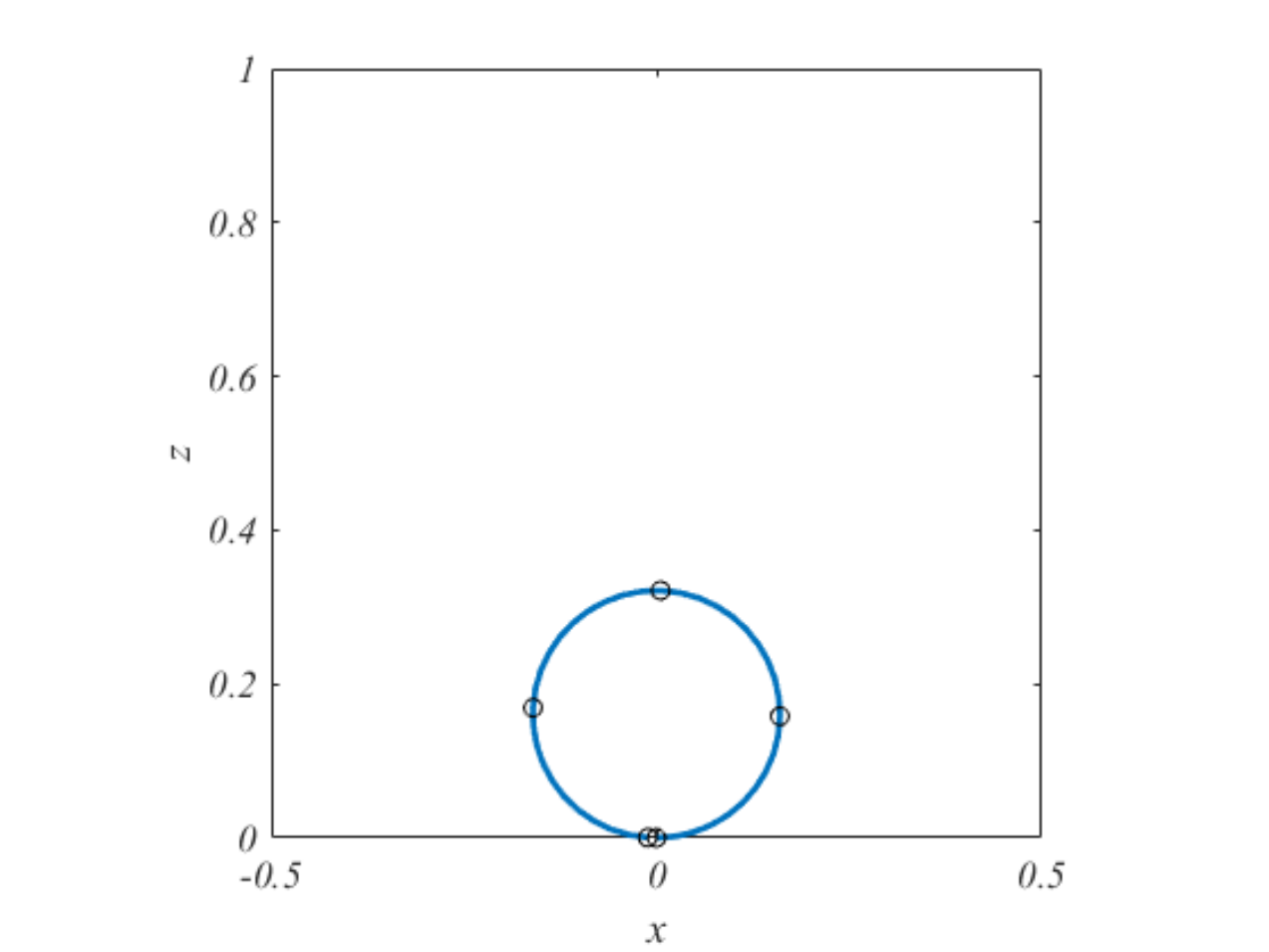}
\caption{Pure bending with $\beta=1$}\label{fig5}
\end{figure} 

\subsection{Euler's elastica}

For the same beam as in the previous example, but under a pinned support on the left end and a roller support on the right, we apply a compressive load of magnitude $21.6$. A small transversal load is applied initially to trigger the post-buckling behaviour. This load of magnitude 21.6, with $\beta=0$, is predicted by the classical theory to make the beam ends meet exactly. Figure \ref{fig6} shows that this is exactly the case according to the finite element implementation, where 6 finite elements have been used. Because of the variable curvature of the elastica, the influence of the strain gradient bending terms is expected to be significant. Figure \ref{fig7} shows the significant stiffening brought about by the values $\beta=0$, $\beta=0.01$, $\beta=0.05$, $\beta=0.1$, $\beta=0.2$, and $\beta=1$, with the same applied load.

 \begin{figure}[H]
	\centering
\includegraphics[scale=0.8, trim=0.in 0in 0in 0.0in, clip]{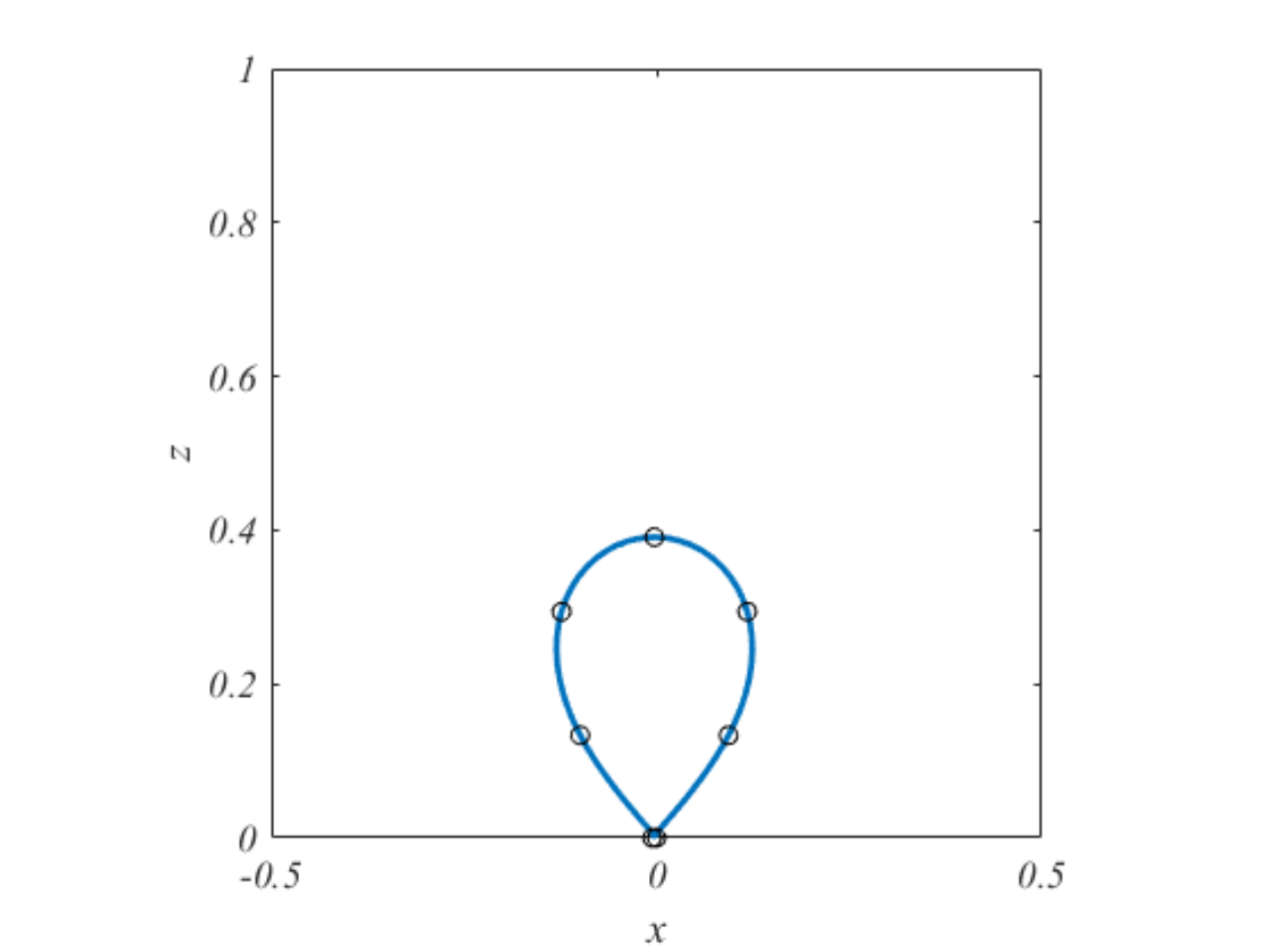}
\caption{The elastica with $\beta=0$}\label{fig6}
\end{figure}

 \begin{figure}[H]
	\centering
\includegraphics[scale=0.8, trim=0.in 0in 0in 0.0in, clip]{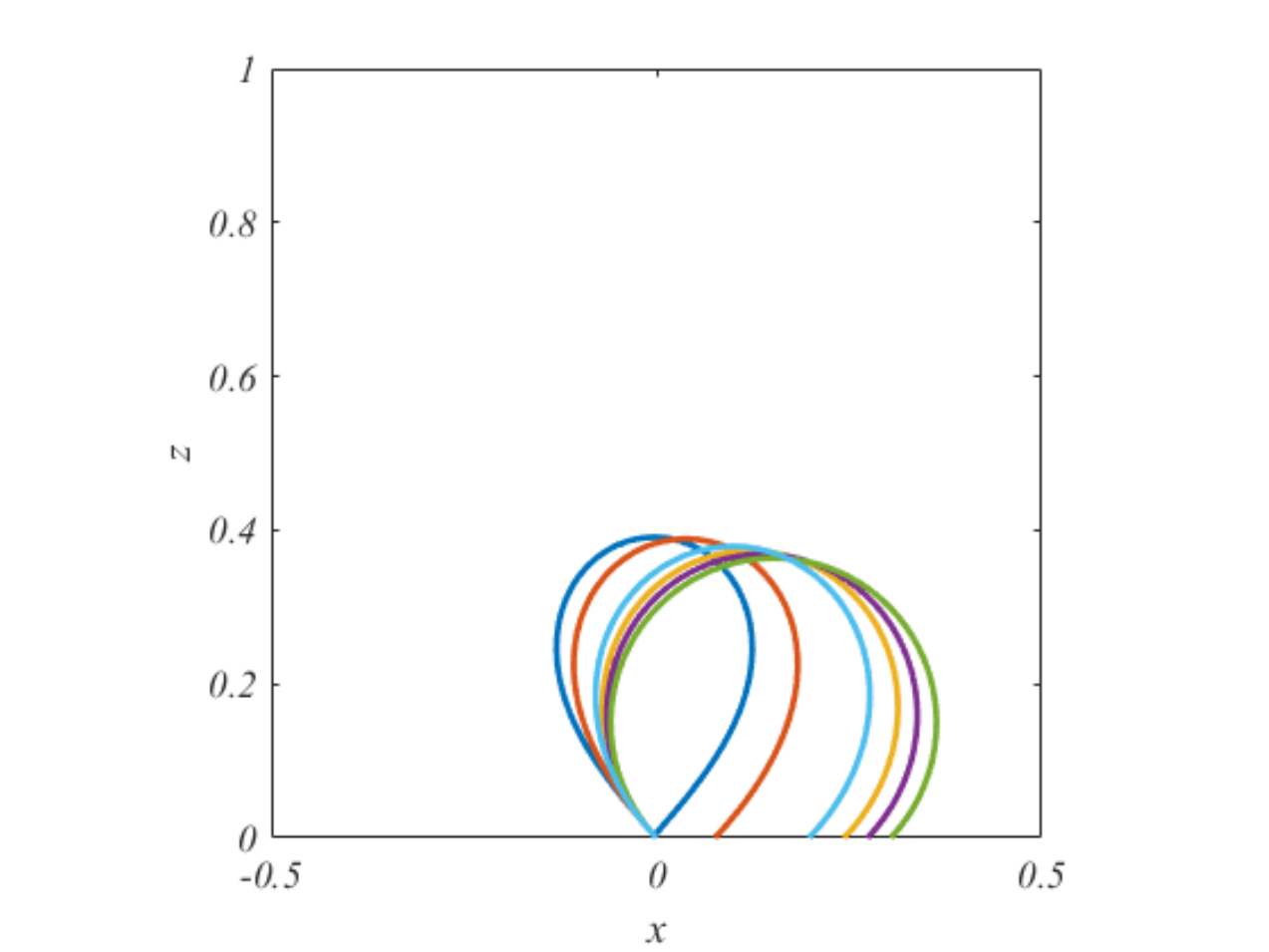}
\caption{The elastica with $\beta=0. 0, 0.01, 0.05, 0.1, 0.2, 1.0$, showing a correspondingly increasing lack of closure and a rounding of the shape}\label{fig7}
\end{figure}

\subsection{Cantilever bending}

Using the same beam properties, we illustrate the large-deflection behaviour of a cantilever, as in Section \ref{sec:linear}, including the geometrically nonlinear effects. Notice the significant horizontal displacement of the tip, not captured by the linear theory. A concentrated upward force of magnitud $5$ is applied at the tip. The corresponding deflection plot for $\beta=0$ is shown in Figure \ref{fig11}. Figure \ref{fig12} shows the deformed beam for the values $\beta=0, 0.1, 1$.

 \begin{figure}[H]
	\centering
\includegraphics[scale=0.8, trim=0.in 0in 0in 0.0in, clip]{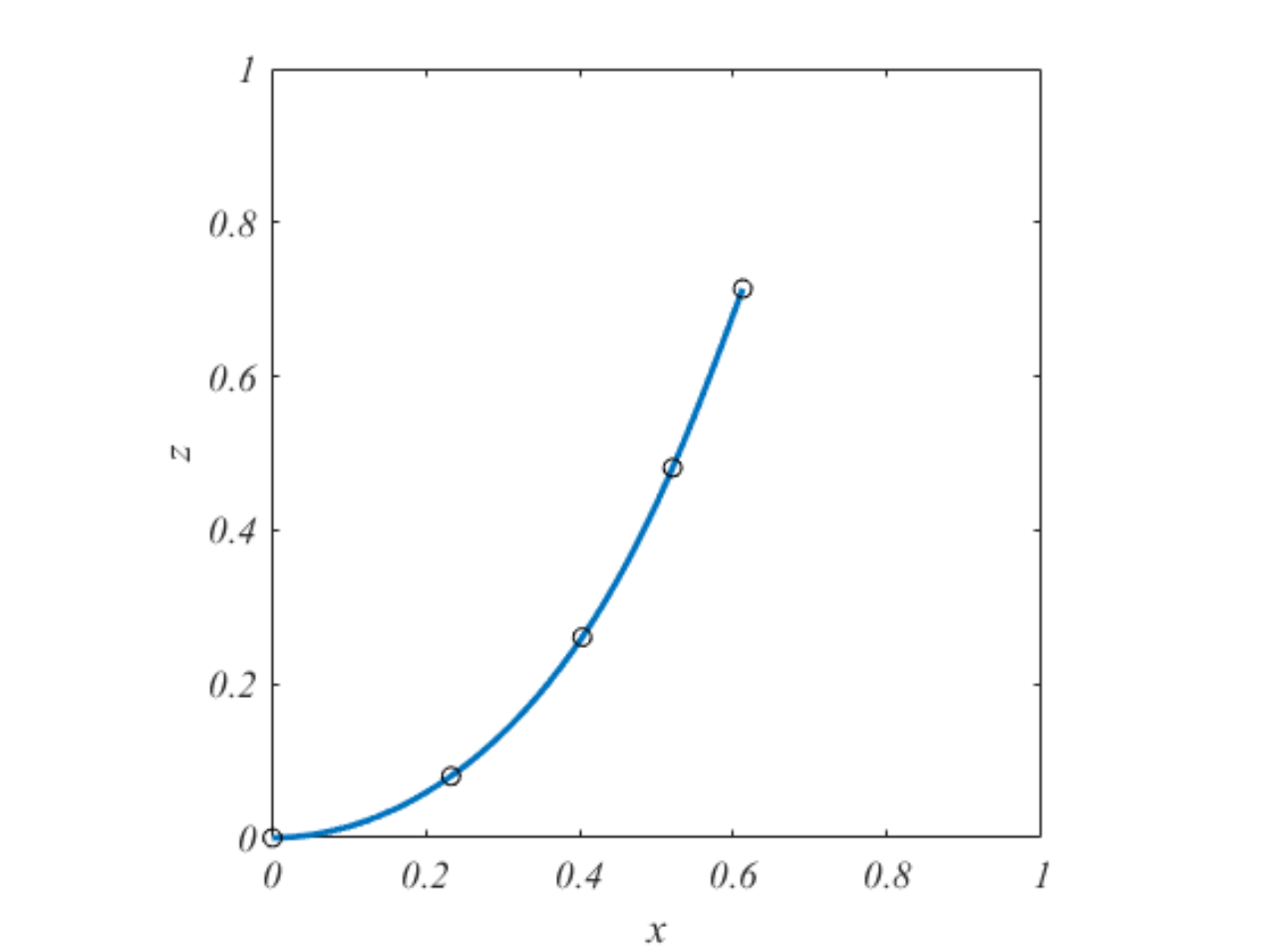}
\caption{Cantilever with tip load $P=5$ and $\beta=0$}\label{fig11}
\end{figure} 

 \begin{figure}[H]
	\centering
\includegraphics[scale=0.8, trim=0.in 0in 0in 0.0in, clip]{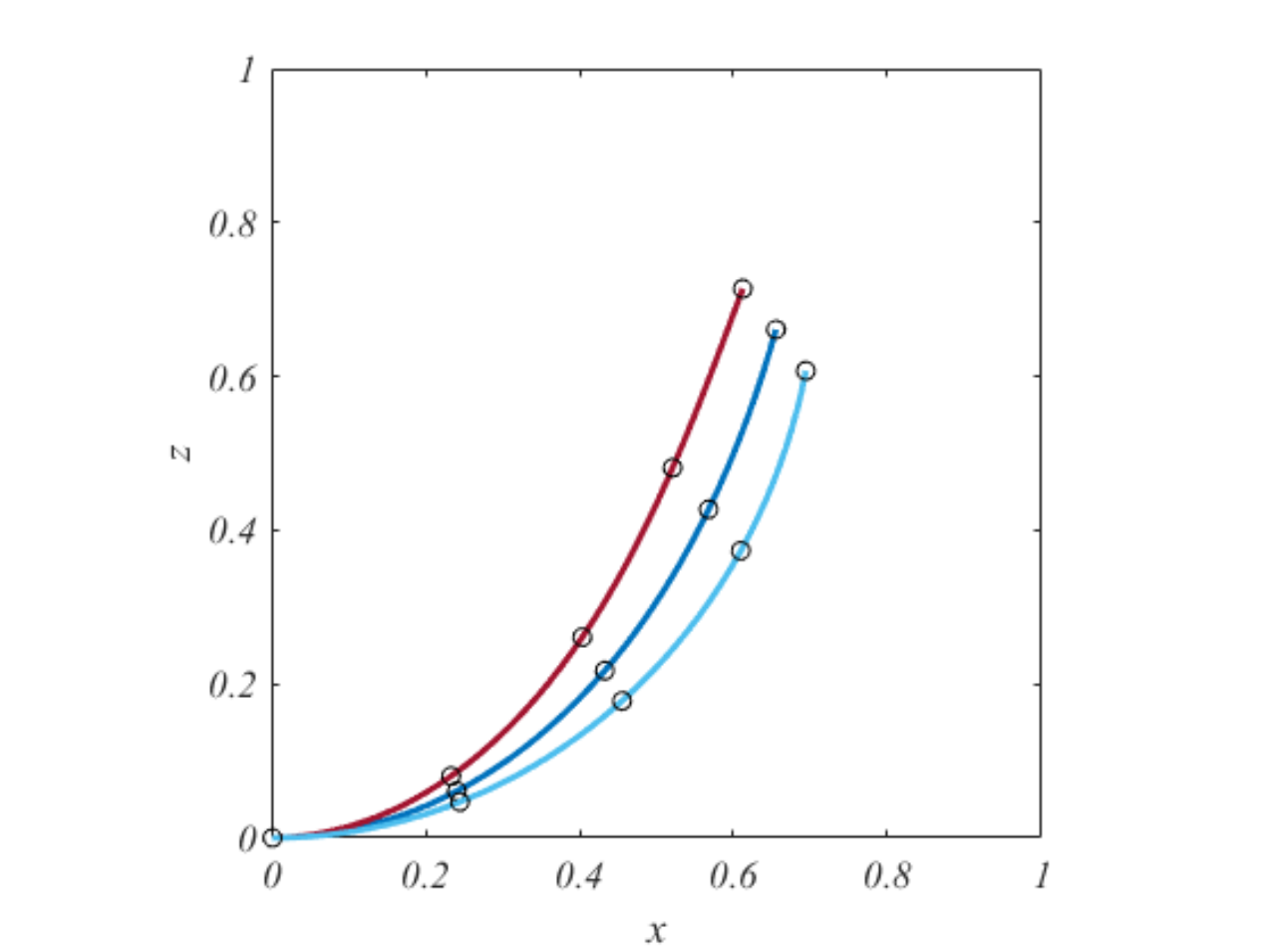}
\caption{Cantilever with $\beta=0, 0.1, 1.0$, showing correspondingly stiffer responses}\label{fig12}
\end{figure}

\subsection{Magnetized bodies}

Perhaps the simplest way to deal with a deformable magnetized body with a fixed (remanent) magnetization field ${\bf B}^r$ when under the action of an externally applied field ${\bf B}^a$, is to express their interaction in terms of a body force per unit volume
\begin{equation} \label{eq100}
{\bf f}=\frac{1}{\mu_0} {\bf B}^r \cdot \nabla {\bf B}^a
\end{equation}
and a body couple per unit volume
\begin{equation} \label{eq101}
{\bf m}=\frac{1}{\mu_0} {\bf B}^r \times {\bf B}^a.
\end{equation}
In these formulas,\footnote{See \cite{brown}, p 54.} $\mu_0$ is a constant (the permeability of vacuum), and $\nabla$ denotes the spatial gradient operator.

Consider a beam with a remanent magnetization given by a constant vector ${\bf B}_0^r$ per unit length of the straight reference configuration.  Let this vector be directed along the beam axis. Upon deformation, the remanent magnetization per unit {\it undeformed} length is given by
\begin{equation} \label{eq102}
{\bf B}^r=\frac{ B_0^r}{\sqrt{1+2e}}\left( (1+u') {\bf i} +w' {\bf k}\right).
\end{equation}
This vector field is, of course, tangential to the deformed axis. Assuming, for specificity, that the external magnetic field ${\bf B}^a=B^a_x {\bf i}+B^a_z {\bf k}$  is spatially constant, the magnetic body force vanishes, while the magnetic body couple per unit undeformed length is given by
\begin{equation} \label{eq103}
{\bf m}=\frac{1}{\mu_0} {\bf B}^r \times {\bf B}^a = -\frac{1}{\mu_0} \frac{B_0^r}{\sqrt{1+2e}}\left( B^a_z (1+u')- B^a_x w'\right) {\bf j}.
\end{equation}
Noting that in Figure \ref{fig:rotation} the unit vector $\bf j$ points into the plane of the page, the counterclockwise version of this couple should ignore the minus sign at the front of the formula.

The variation of Equation (\ref{eq15a}) yields
\begin{equation} \label{eq104}
\delta\theta=\frac{1}{1+2e} \left((1+u') \delta w'- w' \delta u'\right).
\end{equation}
We conclude, therefore, that the external virtual work of the magnetic couples is obtained as
\begin{equation} \label{eq105}
EVW_m=\frac{B_0^r}{\mu_0} \int\limits_0^L \frac{1}{(1+2e)^{3/2}}\left( B^a_z (1+u')- B^a_x w'\right)  \left((1+u') \delta w'- w' \delta u'\right)\,dx
\end{equation}

As defined, the product ${\hat m}=B_0^r B^a/\mu_0$ has the units of a couple per unit length. For the purpose of our numerical examples, we will input the values of $\hat m$ and of the angle $\phi$ measured counterclockwise from $x$ to the vector ${\bf B}^a$, so that $B^a_x=B^a \cos \phi$ and $B^a_z=B^a \sin \phi$. Figure \ref{fig14} shows, for the same cantilever beam as in the previous examples, the deflected beam for $\beta=0$ and ${\hat m}=10$ with $\phi$ varying from $0$ to $90$ degrees, in steps of $15$ degrees. Figure \ref{fig15} shows the corresponding results with $\beta=0.2$.

 \begin{figure}[H]
	\centering
\includegraphics[scale=0.8, trim=0.in 0in 0in 0.0in, clip]{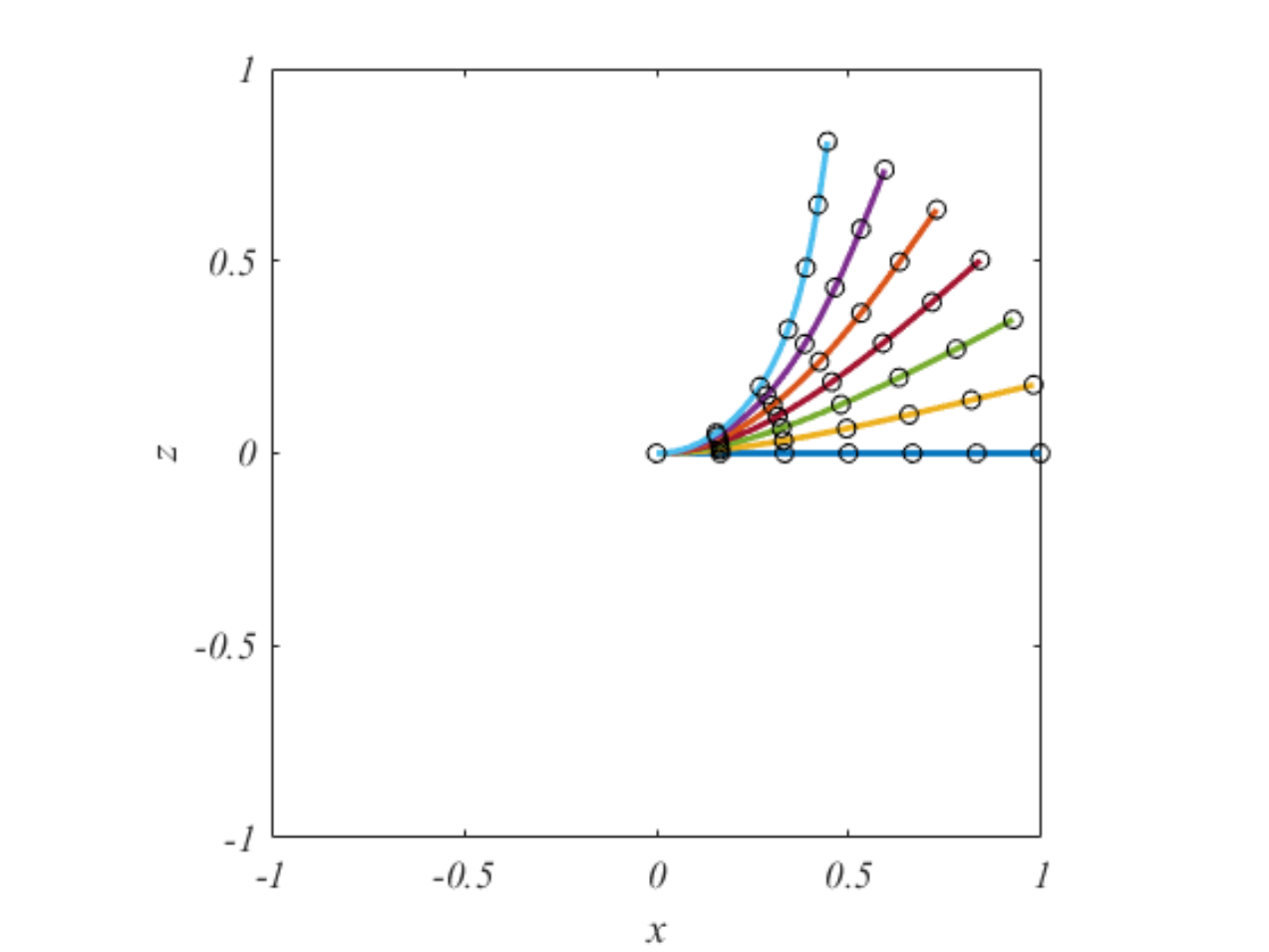}
\caption{Cantilever with $\beta=0$, ${\hat m}=10$ and $\phi=0^o,...,90^o$, showing a correspondingly increasing deflection}\label{fig14}
\end{figure} 

 \begin{figure}[H]
	\centering
\includegraphics[scale=0.8, trim=0.in 0in 0in 0.0in, clip]{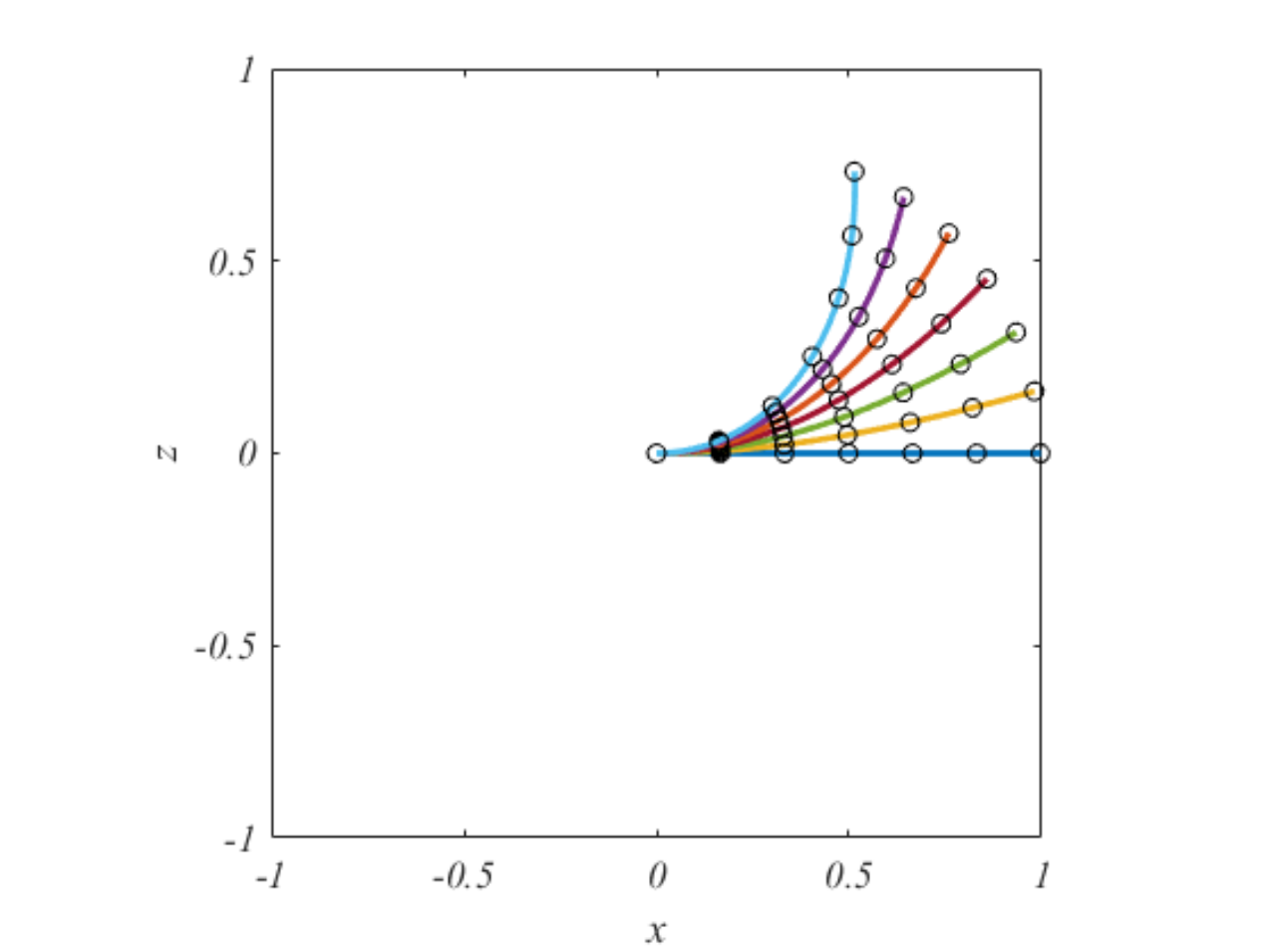}
\caption{Cantilever with $\beta=0.2$, ${\hat m}=10$ and $\phi=0^o,...,90^o$, systematically stiffer than their counterparts in Figure \ref{fig14}}\label{fig15}
\end{figure} 

\subsection{Multiple solutions}

Because the field equations (and their finite-element approximations) are highly nonlinear, we expect multiple solutions to possibly exist for identical loading conditions. The mathematical analysis of the stability of these solutions and of the appearance of bifurcations \cite{dehrouyeh} is beyond the scope of this paper. Numerical procedures in themselves cannot provide a definitive assurance in regard to these issues. The eventual convergence of an iterative procedure depends on the starting guess. Different guesses may lead to no convergence at all, or possibly to different converged solutions, all of them valid. In addition to this factor, the numerical equations being solved depend on the fineness of the finite element mesh. Coarse meshes may unduly rigidify the structure and lead to non-physical solutions. Nevertheless, the fact that different initial guesses to kickstart the Newton-Raphson iterative procedure may lead to different solutions, and that the solutions arrived at are numerically stable, may be a good indication that physically different stable solutions can be observed in carefully conducted experiments.

As an illustration of a non-trivial case of multiple solutions, we return to our magnetized cantilever. Using a relatively fine mesh of 12 equal elements, we will cover the domain of a full rotation of the applied magnetic field, starting at $\phi=0$ in increments of $15$ degrees. We will traverse the domain first in the counterclockwise and then in the clockwise direction. At each step we will use the solution of the previous step as our starting guess. With $\beta = 0$, the results of the first and the second strategies are reproduced in Figures \ref{fig16} and \ref{fig17}, respectively. In this particular example, we observe that the solutions for $-75^o \le \phi \le +75^o$ coincide for both strategies. For each of the remaining angles considered there are two distinct solutions. For angles $105^o \le \phi \le 255^o$, the two solutions appear to be stable. The outliers, clearly distinguishable in each figure, happen to be $\phi=90^o$ for the clockwise sweep, and $\phi=-90^o$ for the counterclockwise one. The Newton-Raphson procedure failed to converge for these two angles in their respective sweeps. It would appear that the solution bifurcates at about $\phi=\pm 90^o$. No general conclusions can be drawn from a numerical example. The enormous power of computational mechanics is also a manifestation of its weakness.

 \begin{figure}[H]
	\centering
\includegraphics[scale=0.8, trim=0.in 0in 0in 0.0in, clip]{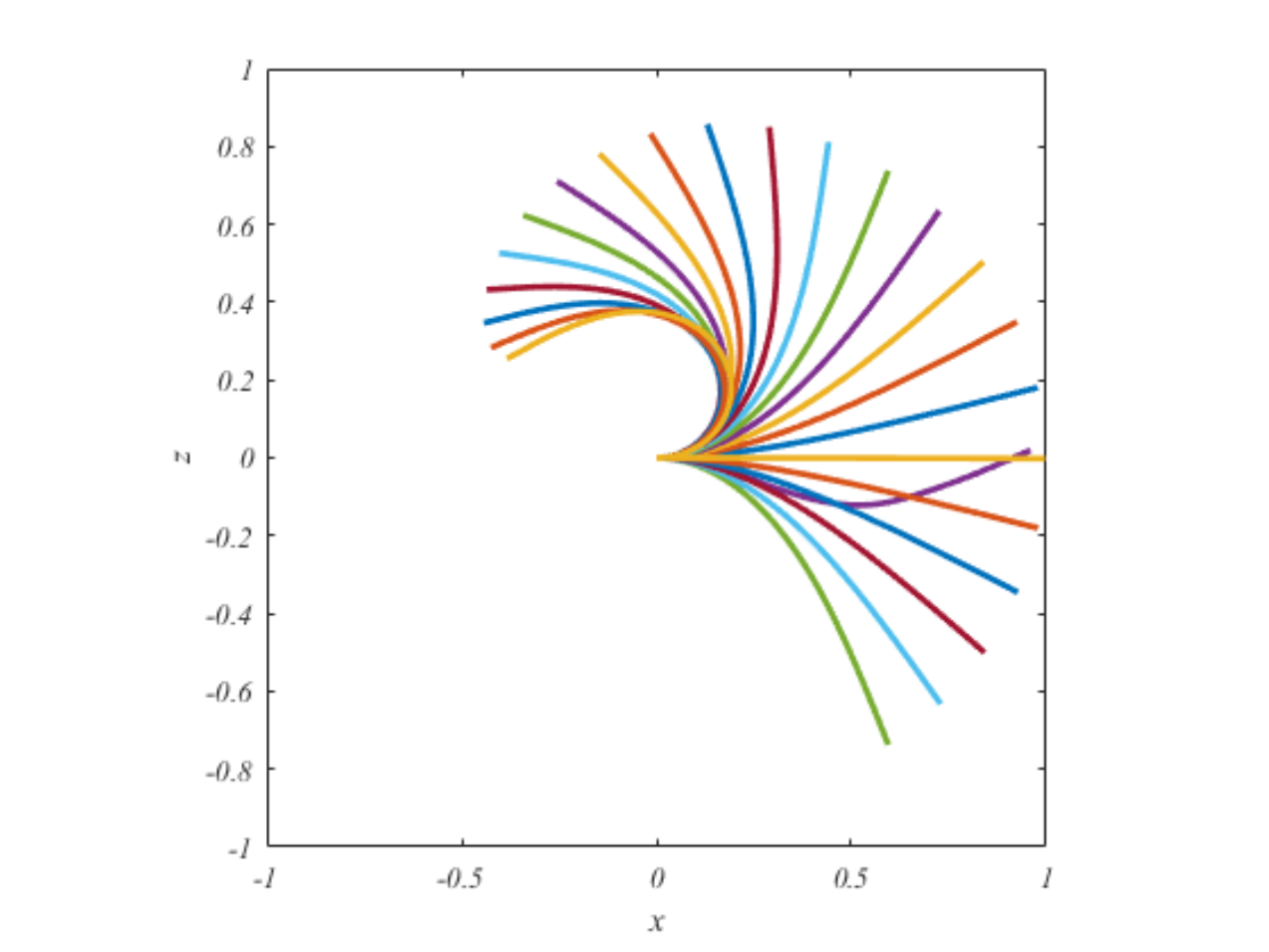}
\caption{Cantilever with $\beta=0$, ${\hat m}=10$ and $\phi$ sweeping a full circle counterclockwise at $15^o$ intervals}\label{fig16}
\end{figure}

 \begin{figure}[H]
	\centering
\includegraphics[scale=0.8, trim=0.in 0in 0in 0.0in, clip]{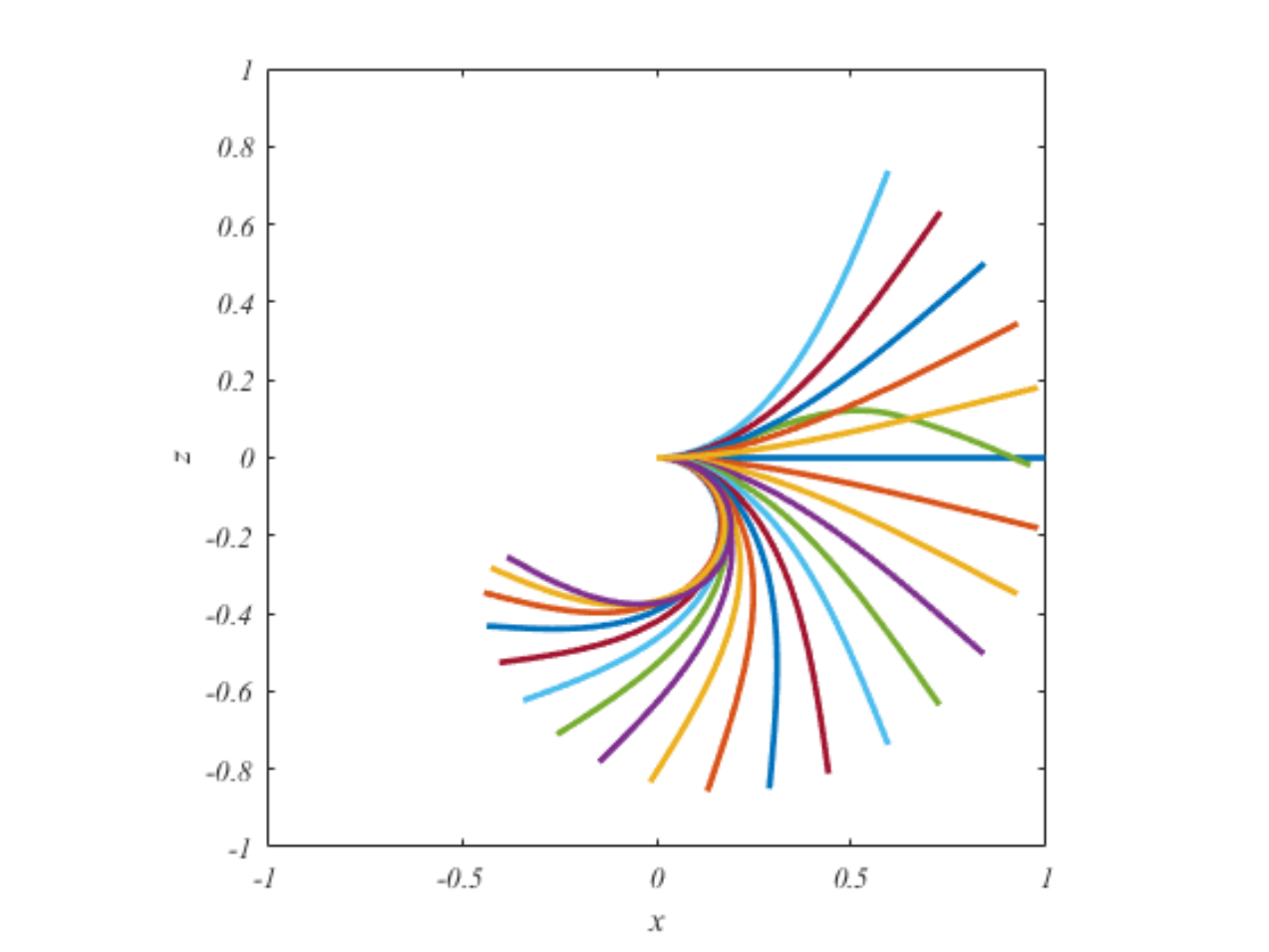}
\caption{Cantilever with $\beta=0$, ${\hat m}=10$ and $\phi$ sweeping a full circle clockwise at $15^o$ intervals}\label{fig17}
\end{figure}

\end{document}